# Valleytronics in bulk MoS$_2$ by optical control of parity and time symmetries

Igor Tyulnev[1], Álvaro Jiménez-Galán[2,3], Julita Poborska[1], Lenard Vamos[1], Rui F. Silva[4], Philip St. J. Russell[5,6], Francesco Tani[5], Olga Smirnova[2,7], Misha Ivanov[2,8,9], Jens Biegert[1,10,*]

[1] ICFO - Institut de Ciencies Fotoniques, The Barcelona Institute of Science and Technology, 08860 Castelldefels (Barcelona), Spain
[2] Max-Born-Institut, Max-Born-Str. 2A, 12489 Berlin, Germany
[3] Joint Attosecond Science Lab, National Research Council of Canada and University of Ottawa, Ottawa, ON K1A 5N2, Canada
[4] Instituto de Ciencia de Materiales de Madrid (ICMM), Consejo Superior de Investigaciones Científicas (CSIC), 28049 Madrid, Spain
[5] Max-Planck Institute for Science of Light, Staudtstraße 2, 91058 Erlangen, Germany
[6] Department of Physics, Friedrich-Alexander-Universität, Staudtstraße 2, 91058 Erlangen, Germany
[7] Technische Universität Berlin, Hardenbergstr. 33A, 10623 Berlin, Germany
[8] Institute für Physik, Humboldt-Universität zu Berlin, Newtonstr. 15, 12489 Berlin, Germany
[9] Department of Physics, Imperial College London, SW7 2AZ London, United Kingdom
[10] ICREA, Pg. Lluís Companys 23, 08010 Barcelona, Spain
*Correspondence to: jens.biegert@icfo.eu

**The valley degree of freedom of electrons in materials promises routes toward energy-efficient information storage with enticing prospects towards quantum information processing. Current challenges in utilizing valley polarization are symmetry conditions that require monolayer structures or specific material engineering, non-resonant optical control to avoid energy dissipation, and the ability to switch valley polarization at optical speed. We demonstrate all-optical and non-resonant control over valley polarization using bulk MoS$_2$, a centrosymmetric material with zero Berry curvature at the valleys. Our universal method utilizes spin-angular momentum-shaped tri-foil optical control pulses to switch the material's electronic topology to induce valley polarization by transiently breaking time and space inversion symmetry through a simple phase rotation. The dependence of the generation of the second harmonic of an optical probe pulse on the phase rotation directly demonstrates the efficacy of valley polarization. It shows that direct optical control over the valley degree of freedom is not limited to monolayer structures. Instead, it is possible for systems with an arbitrary number of layers and bulk materials. Universal and non-resonant valley control at optical speeds unlocks the possibility of engineering efficient, multi-material valleytronic devices operating on quantum coherent timescales.**

Lightwave electronics[1–6] envisions optical control over charge carrier dynamics in materials on a sub-cycle timescale, aiming at information processing at unprecedented petahertz rates. In addition to increased speed, a core requirement is an ability to encode information into electronic degrees of freedom to realize classical bits, with material-intrinsic quantum correlations providing an additional path toward lightwave qubit applications[7,8]. An enticing prospect to encode, process, and store quantum information provides the valley pseudospin[9–12], associated with local extrema in the electronic bands of the first Brillouin zone in transition metal dichalcogenide (TMDC) semiconductors. Such materials exhibit two energy-degenerate valleys at the crystal momenta's K and K' high-symmetry points. In the monolayer form, where the inversion symmetry is broken, the K and K' valleys can be addressed selectively using right- and left-handed circularly polarized light, respectively, thanks to an optical valley selection rule that can be traced back to both the orbital character of the bands and their Berry curvature[11,13–17]. However, in multi-layer and bulk TMDCs, such optical valley selection is not possible due to the opposing orientation of layers of transition metal and chalcogen atoms that occurs in their most stable configuration.

Figure 1a shows the momentum-resolved electron distribution that results from the interaction of bilayer $MoS_2$ and left-hand circularly-polarized light using time-dependent first-principles calculations. The density plot projection shows that all high symmetry points of the hexagonal first Brillouin zone are equally populated since both valleys, K and K', are indistinguishable. Reduction of the thickness of the TMDC to the monolayer remedies this issue by breaking space inversion symmetry. A circularly polarized resonant optical field provides the additional time inversion symmetry breaking, thus, resulting in valley pseudospin selectivity; see Fig. 1b. This optical valley selectivity has been achieved in several TMDC monolayers[13,18].

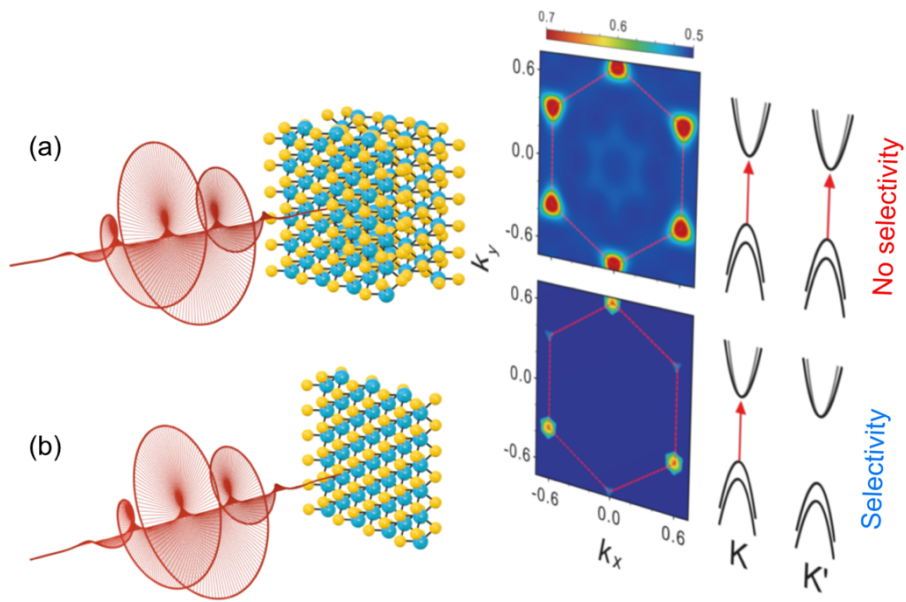

Figure 1. Theoretical results from ab-initio density functional theory show valley polarization prospects in 2H-$MoS_2$. (a) Left-hand circularly polarized gap-resonant excitation for thicknesses beyond the monolayer does not result in valley-selective excitation. The k-space image and bandgap illustration show the transfer of population at K and K' from the valence band (VB) to the conduction band (CB). (b) Identical excitation in the monolayer leads to valley selective resonant population transfer here at the K point.

These results encouraged manifold efforts to overcome the limitations imposed by monolayer materials by manipulating the dielectric environment, e.g., through strain engineering[19], different stacking configurations[20], or through an external energy bias between the layers[21,22]. Interlayer excitons[23–25] in type II aligned TMDC heterostructures are investigated to overcome inherent limitations imposed by long-range Coulomb effects leading to dephasing due to intervalley exchange interaction[26,27] and spin relaxation[28]. Yet, significant obstacles to realizing the true prospects of light wave valleytronics are i) the gap-resonant excitation that results in material-specific valley switching rates and optical setups and ii) the need for inversion symmetry breaking by dedicated material engineering. Worse, when exfoliated to the monolayer, most materials with interesting inherent quantum properties are severely limited in available quality, size, or environmental stability. A material-agnostic optical control scheme would permit leveraging materials, or a combination of materials, with specific properties that are otherwise not accessible. Reference 6 already hinted at such a possibility when replacing the gap-resonant circularly polarized optical field with a strong optical field in monolayer $MoS_2$.

**Universal valley control with a non-resonant tri-foil optical field – physical mechanism**

Here, we will show that a strong non-resonant tri-foil optical field[29], as shown in Fig. 2a, provides a universal and material-agnostic means to switch and control valley polarization in bulk hexagonal materials optically. The optical field must be SAM-shaped to break *both* time and space-inversion symmetry since the single-layer triangular sub-lattice of hexagonal materials is three-fold symmetric. Figure 2a shows results from a time-dependent first-principles simulation of bilayer 2H-$MoS_2$ with such a strong tri-foil field applied. The k-space projected electronic density clearly shows valley polarization despite the 180-degree rotation between two layers.

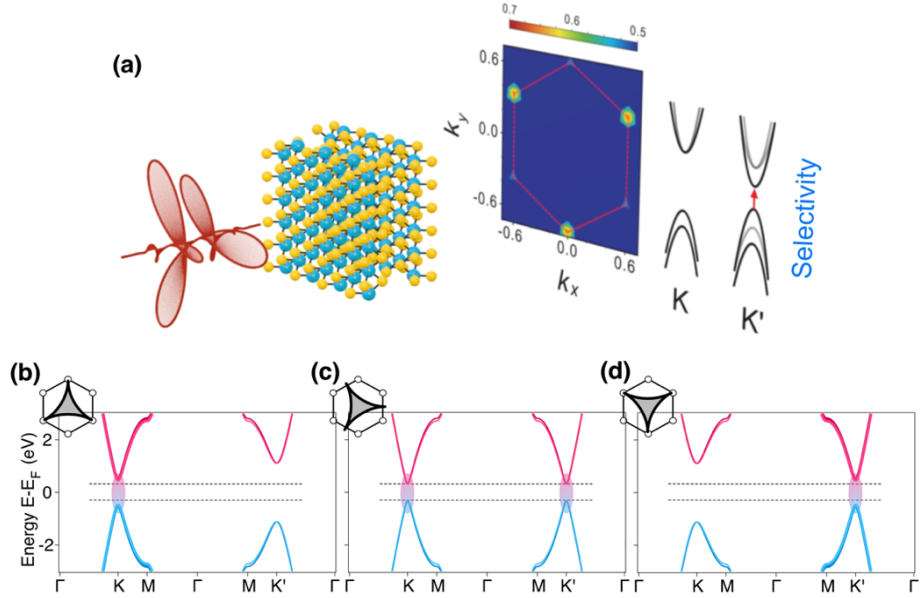

*Figure 2. Valley polarization in bulk 2H-$MoS_2$ with a strong tri-foil field. (a) Simulated interaction between a strong tri-foil field and 2H-$MoS_2$, calculated from first principles, showing strong polarization at the K valley (color code). Valley polarization switches when the tri-foil field is rotated by 60 degrees. (b-d) The physical mechanism of the effect is related to the laser-induced effective band structure, illustrated here using a four-band tight-binding model of AA-stacked gapped graphene modified by the tri-foil field. The parameters of the model are adjusted to mimic two layers of 2H-$MoS_2$. The strong tri-foil field induces complex second neighbor hopping that breaks time-reversal symmetry and lifts the valley degeneracy (see SI). The values of the laser-modified hopping depend on the orientation of the tri-foil field with respect to the lattice, as shown in the insets. Valley polarization switches between K (b) and K' (d) with a simple 60-degree optical rotation of the tri-foil field. A 30-degree rotation results in equal valley polarization between K and K' (c).*

In this case, the mechanism for valley polarization is fundamentally different from that enabled by the optical valley selection rules. In this case, this effect is purely driven by symmetries. On the one hand, the space inversion symmetry is broken in the composite laser-dressed crystal thanks to the matching symmetry between the tri-foil field and the layer sub-lattice; see the insets of Fig. 2. On the other hand, that same symmetry can be shown to induce complex second neighbor hopping[6] similar to those present in the topological model of Haldane, which break the time-reversal symmetry and lift the valley degeneracy. While the latter effect was theoretically predicted for a broken-inversion symmetric monolayer structure, Fig. 2a shows that the same applies to the centrosymmetric bi-layer case. To illustrate the physical mechanism, we complement numerical simulations of Fig. 2a with the effective band structure of a dressed gapped graphene system (Figures 2b-d). We analytically calculate the laser-modified hopping using our experiment's laser parameters and the field-free bilayer 2H-$MoS_2$. Visible is the minimal gap switching between K and K' points for 60-degree rotation of the tri-foil field (Fig. 2b,d). Figure 2c shows the case in which the orientation of the tri-foil field affects both valleys equally, that is, the induced second neighbor hoppings are real.

**Non-resonant tri-foil optical control with a tri-foil field: experiment**

For the experiment, we synthesize the tri-foil pulses from our mid-IR OPCPA system[30] in a two-color Mach Zehnder interferometer by combining a circularly-polarized field with frequency $\omega$ with its circularly-polarized second harmonic $2\omega$ rotating in the opposite direction. The combined field forms a tri-foil SAM-shaped optical pulse, and the field's orientation is controlled, on a sub-laser-cycle timescale, through the phase delay between the two optical fields at $\omega$ and $2\omega$.

The laser system delivers carrier-envelope-phase (CEP) stable, 3.2-micron, 103-fs pulses at a repetition rate of 160 kHz. A small energy part of this output enters the Mach-Zehnder interferometer, generating the second harmonic in one of the arms. Both arms contain half and quarter waveplates to adjust the relative field amplitudes and polarization states of the fundamental ($\omega$) and second harmonic ($2\omega$). A suitable phase delay and amplitude ratio of 3:2 ($\omega:2\omega$) for opposite helicity fields result in the generation of the tri-fold SAM structure upon the coherent combination of the two arms. The parameters of the Mach-Zehnder interferometer are fine-tuned and checked with a control experiment to ensure the generation of the tri-foil field. To this end, we use chiral high harmonic spectroscopy (HHS)[31–39] in GaSe to tune the Mach-Zehnder from linear to circular and to bi-circular to confirm the synthesis of the tri-foil field. The measured symmetry response[32] in the high harmonic spectrum reveals six-fold modulation for linear, no modulation for circular, and three-fold modulation for the bi-circular optical fields. This measurement directly confirms the tri-foil character of the SAM-synthesized bi-circular optical field.

We now consider the experimental validation of strong-field non-resonant valley control in a centrosymmetric material. To test this prospect, we apply the tri-foil field to a free-standing 50-micron thick sample of bulk 2H-MoS$_2$. Figure 3 shows results from a transmission measurement in a free-standing 2H-MoS$_2$ sample for a tri-foil control field at normal incidence and with a field amplitude of $0.032 \pm 0.004$ V/Å.

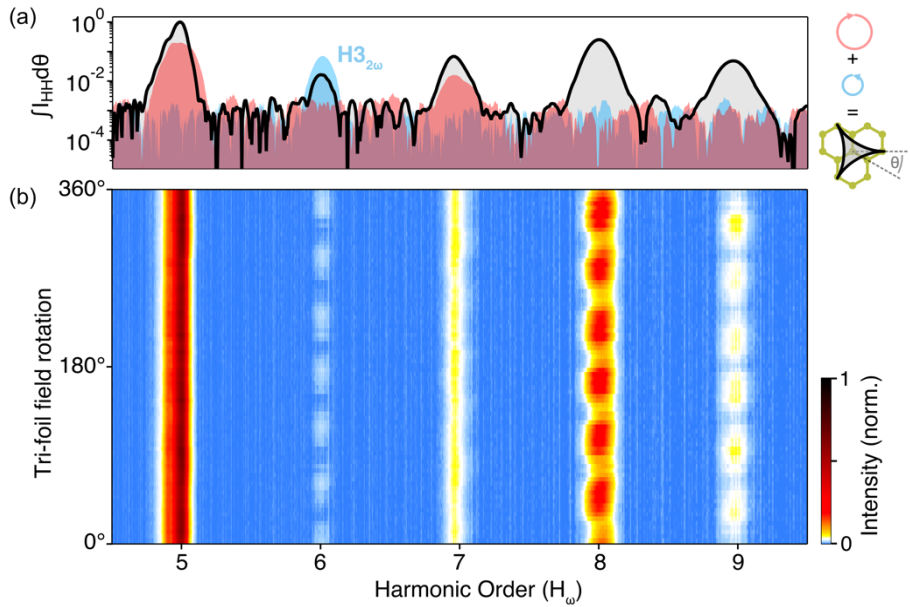

*Figure 3. High harmonic spectroscopy (HHS) of free-standing 50-μm-thick MoS$_2$. (a) Spectra, integrated over 360-degree rotation of the control/pump field. The right-hand circularly polarized 3.2-μm field ($\omega$) generates odd-order harmonics (H5, H7) as expected for a centrosymmetric material. The weaker second harmonic field ($2\omega$) only yields its third harmonic (H3$_{2\omega}$), whose frequency coincides with H6 of the fundamental. Due to energy and spin angular momentum conservation, the tri-foil field yields harmonics H($3N \pm 1$). The symmetry forbidden H6 arises due to*

*generation at the material interface and not from the bulk. (b) Rotation-resolved HHS for the tri-foil field. The nonlinear intensity scale shows a six-fold modulation, and the harmonic order on the horizontal axis is for the 3.2-micron field.*

Figure 3(a) shows the high harmonic spectra for a right-handed circular field at 3.2 microns (red-filled curve), the left-handed circular field at 1.6 microns (blue-filled curve), and the combination of both (tri-foil field, black curve) integrated over a full rotation. The right-handed circular 3.2-micron field generates only odd harmonic orders H5 and H7, as expected from centrosymmetric bulk 2H-MoS$_2$. Similarly, the left-handed circularly polarized second harmonic field shows only its third harmonic, H6 of the fundamental field. Since the second harmonic field amplitude is only 2/3rds of the field amplitude of the fundamental, higher harmonics are not observed. In the monolayer limit, with symmetry space group P-6m2 due to the broken spatial symmetry, 3Nth harmonics are symmetry-forbidden in circularly-polarized fields. In stark contrast to the individual field cases, the tri-foil field yields even harmonic orders. Harmonic order H8 emerges as the strongest signature, 2.5 dB above the noise floor. Odd harmonic orders H5 and H7 increase by a factor 4.6 on average, while H6 is suppressed by a comparable factor of 4.3. The dramatically different behavior of H6 is a consequence of the breaking of spatial inversion symmetry. The resulting C$_3$ symmetry leads to the $H(3N \pm 1)$ selection rule for the high harmonic generation with harmonic orders H(3N) are suppressed (N is a positive integer). The non-zero signal in the case of H9 is attributed to breaking the tri-foil field symmetry since it lies at the harmonic cut-off where only the most intense parts of the synthesized light wave contribute to the harmonic generation instead of the entire tri-foil structure. Figure 3(b) shows how most harmonics exhibit a precise six-fold symmetry as a function of tri-foil field rotation, thus, confirming the purity of the 2H phase of the material.

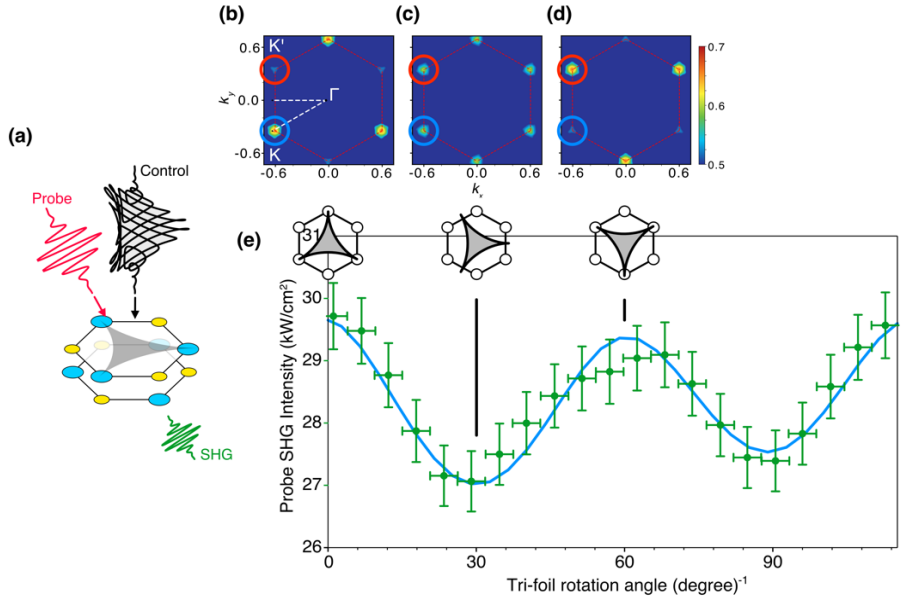

*Figure 4. Probing valley polarization in bulk MoS$_2$. (a) The tri-foil control field impinges at normal incidence to induce and control valley polarization in the material. The efficacy of valley polarization and symmetry breaking is measured by generating the second harmonic (SHG) of an 800 nm probe pulse in a background-free geometry. Without the tri-foil control field, no SHG is observed from the bulk phase. (b)-(d) shows calculated k-space projections of the expected valley polarization. The tri-foil field is shown in real space as insets. (e) The measured SHG intensity (green dots) agrees well with the predicted SHG intensity (blue line) as a function of tri-foil field rotation.*

Next, we probe the generated valley polarization in bulk 2H-MoS$_2$. A weak 800-nm field with a field strength of $0.012 \pm 0.002$ V/Å is used to examine the change in the electronic structure and breaking of time and space inversion symmetry as a function of the tri-foil field rotation angle. The tri-foil field impinges at normal incidence, while the 800-nm probe field interrogates the material non-collinearly in s-polarization, thus providing a background-free measurement. Without a tri-foil field, no second-harmonic generation (SHG) of the probe is observed from the bulk phase to the centrosymmetric nature of multi-layer 2H-MoS$_2$. Figure 4 shows the geometry together with calculated and experimental results. With the tri-foil field present and at maximum pump-probe overlap, the emergence of the second harmonic of the probe field at 400 nm is a clear signature of symmetry breaking. The functional behavior of the SHG signal (green dots) as a function of tri-foil field rotation is faithfully reproduced by our calculation (blue line). Figure 4(b)-(d) shows how valley polarization switches between the K (blue circle) to K' points (red circle). In agreement with the theory, the SHG signal in Fig. 4(e) shows proportionality to the degree of valley polarization, reaching a maximum when the pump polarization is oriented along the lattice, thus confirming the strong-field-induced valley control. It reveals that the delay changes the electron population, as measured by the SHG of the probe signal, from the K valley, at zero delay, to an equal distribution between K and K' valleys for a delay of 30 degrees and to the K' valley at a delay of 60 degrees. The switching repeats with the period of the $2\omega$ part of the tri-foil field. The fact that the SHG signal in Fig.4e is non-zero for 30 degrees and 90 degrees is due to the anisotropy of the tri-foil structure, which breaks the symmetry even in the absence of valley polarization. However, this symmetry breaking is minor compared to that produced by valley polarization (see Fig. 4e).

## Summary

We have shown that a strong SAM-shaped optical field, whose symmetry matches the sublattice of the material, achieves high contrast valley polarization even in a bulk material with net-zero Berry curvature. Here we demonstrate optical control over valley polarization in bulk 2H-MoS$_2$, a centrosymmetric material without Berry curvature. Our universal and material-agnostic method allows accessing a wide variety of enticing materials, e.g., quantum solids with intrinsic entanglement, topological phases, or spin textures, which are difficult to produce with adequate size as monolayers. Moreover, sub-cycle control over the material's electronic topology to induce valley polarization by transiently breaking time and space inversion symmetry through a simple phase rotation may offer an entirely new approach to ultrafast lightwave valleytronics for low-loss information processing in solids. Combining such capability with quantum materials with intrinsic correlations may further provide new avenues for strong field quantum information processing and producing massively entangled states on ultrafast time scales.

## Acknowledgment

J.B. acknowledges financial support from the European Research Council for ERC Advanced Grant "TRANSFORMER" (788218), ERC Proof of Concept Grant "miniX" (840010), FET-OPEN "PETACom" (829153), FET-OPEN "OPTOlogic" (899794), FET-OPEN "TwistedNano" (101046424), Laserlab-Europe (871124), MINECO for Plan Nacional PID2020–112664 GB-I00; AGAUR for 2017 SGR 1639, MINECO for "Severo Ochoa" (CEX2019-000910-S), Fundació Cellex Barcelona, the CERCA Programme/Generalitat de Catalunya, and the Alexander von Humboldt Foundation for the Friedrich Wilhelm Bessel Prize. I.T. and J.B. acknowledge support from Marie Skłodowska-Curie ITN "smart-X" (860553). A.J.G. acknowledges support from the EU Marie Skłodowska-Curie Global Fellowship (101028938) and the Joint Center for Extreme Photonics. R.F.S. acknowledges support from the fellowship LCF/BQ/PR21/11840008 from "La Caixa" Foundation (ID 100010434). M.I acknowledges support from FET-OPEN "OPTOlogic" (899794). We thank U. Elu, M. Enders and L. Maidment for assistance.